\input epsf
\documentstyle[prd,floats,aps]{revtex}

\begin{document}

\vskip1cm
\centerline{\LARGE \bf Mixmaster quantum cosmology}
\centerline{\LARGE \bf in terms of physical dynamics}
\vskip1cm
\centerline{Seth Major and Lee Smolin}
\vskip1cm
\centerline{\it   Center for Gravitational Physics and Geometry}
\centerline{\it The Pennsylvania State University}
 \centerline {\it University Park, Pennsylvania, 16802-6360   U.S.A.}
\centerline{July 5, 1996}
\vskip2cm
\centerline{ \bf Abstract}
\noindent
An approach to quantum cosmology, relying on strengths of
both canonical and path integral formalisms, is applied to
the cosmological model, Bianchi type IX. Physical quantum states
are constructed on the maximal slice of the cosmological history.  A path 
integral is derived which evolves observables off the maximal slice.
This result is compared a path integral propagator derived earlier with 
conventional Faddeev-Poppov gauge fixing.

\section{Introduction}

Previously, we suggested a new approach to
the quantum description of cosmology \cite{SMLS}.   
Relying on the strengths of both canonical and path integral methods, 
this ``composite approach'' to quantization offers a real possibility of
calculating physical observables for cosmological models.
The idea is to define the physical
observables and inner product of the canonical theory
and then use them to construct a path integral which 
evolves these physical observables.  In the 
companion paper, 
we presented a path integral quantization of
the cosmological model Bianchi type IX using  
Faddeev-Poppov techniques familiar from
gauge theory. Here, we carry out this composite approach 
for the model cosmology Bianchi type IX and 
compare our result to that found previously \cite{SMLS}.

One of the great difficulties in the canonical quantization of
gravity is that, to find the physical quantities in a gauge invariant
formalism, it is essentially
necessary to solve the theory.  However, as pointed out
by Vince Moncrief
during a discussion three years ago at Santa Barbara,
there are cases in which we can identify
physical observables and define their algebra.  
For example, in spatially compact cosmological models, we can
define the physical observables to be the metric and extrinsic
curvatures of the maximal slice, on which the trace of the
extrinsic curvature vanishes.  As the metric and extrinsic
curvature evaluated there label solutions uniquely, they
coordinatize the physical phase space.  Moreover, the
Poisson brackets are the naive ones, evaluated on that
surface.  

Alternatively, physical observables can be defined by 
a complete gauge fixing of the theory.  At the classical level,
the two procedures must agree.  However, in the quantum 
theory, it is not obvious that either of these procedures must
lead to a good quantum theory.  Nor is it obvious that they
will lead to the same quantum theory.  Our main goal in this work
is to examine and compare 
the quantum theories that follow from each
of these procedures, for the Bianchi type IX model.  
In each case we construct the path
integral formulation of the quantum theory from the
associated canonical theory.  We compare the resulting
path integrals with each
other as well as with the expressions we previously found
using the Faddeev-Poppov procedure.

We work with the Bianchi type IX cosmological model because
it is the simplest model that is, as far as is known, not solvable
in closed form.  This means that, although the configuration space
is two dimensional, it shares with the real theory the property that
we know of no procedure to explicitly construct the solutions of the
theory.  
The model describes a cosmological family with homogeneous but 
anisotropic spatial slices.  These anisotropies are the two 
dynamical degrees of freedom.  The model has been studied
extensively, especially since
Misner expressed the dynamics as single particle mechanics in
a time dependent potential \cite{MTW,BK}.  A key property, which we shall
exploit, is that all of this model's classical histories 
recollapse \cite{LIN}.  Recently, quantizations of Bianchi type IX 
have been proposed by Kodama \cite{KOD} (see
also \cite{MR}), Grahm \cite{GRA}, and Marolf \cite{DM}.

The paper begins with a quick review of the phase space of this model.
Once the maximal slice is determined we present the physical
quantization and path integral.  The third section offers a comparison
with our earlier work.  The paper finishes with discussion in Section 4. 

We present the derivation in ``geometrized units,'' in which $G = c = 1$.

\section{Summary of the classical theory}

We consider class A Bianchi type IX models which describe
 anisotropic, homogeneous solutions
to the vacuum Einstein equations.\footnote{The classification of 
Bianchi models 
concerns the irreducible
parts of the structure constants of the isometry Lie group.
Writing the structure constants in terms of these pieces, 
$C^I_{JK} = \epsilon_{JKL}S^{LI} + \delta^I_{[J} V_{K]}$, class A models
are those for which $V_I=0$.}
In Misner's chart on the phase space of Bianchi type IX the spatial metric
is given by \cite{MTW}
\begin{equation}
h_{ij} = e^{2\alpha}\left( e^{2\beta} \right)_{ij}
\end{equation}
where $\beta^{ij}$ is diagonal and traceless matrix parameterized by
diag$\left(\beta^+ + \sqrt{3} \beta^- , \beta^+ - \sqrt{3} \beta^-,
-2 \beta^+ \right)$ and $\alpha$ is the scale factor of the cosmology.  
As $\beta^{ij}$ is traceless, the scale factor 
for the cosmology, $\alpha$, is related to the volume through 
$\sqrt{h} = e^{3\alpha}$.
The action for the model Bianchi type IX in this chart is \cite{MTW}
\begin{equation}
I = \int p_+ d\beta^+ + p_- d\beta^- + p_\alpha d\alpha
- \sqrt{ { 3 \pi \over 2}} N e^{-3\alpha} {\cal H} dt
\label{theaction}
\end{equation}
in which $ {\cal H} = {1 \over 2} \left( - p_\alpha ^2 + p_+ ^2 + p_- ^2
+e^{4\alpha} {\cal U} (\beta^\pm) \right)$.\footnote{Note that in
the literature one often finds written 
$V(\beta_\pm ) := {\cal U} (\beta_\pm) +1$.  This is convenient because
$V(\beta_\pm)$ is positive definite, however what is
important to remember is that the actual potential
${ \cal U} (\beta_\pm$) is bounded from below by $-1$.}
The triangularly shaped potential is given by
\begin{eqnarray}
{\cal U}(\beta^\pm) &= &{1 \over 3} e^{-8 \beta^+} 
- {4 \over 3} e^{-2 \beta^+}
\cosh (2 \sqrt{3} \beta^-) 
\nonumber \\
&& +
{2 \over 3} e^{4\beta^+} \left (\cosh (4 \sqrt{3} \beta^-) -1 \right ).
\end{eqnarray}
Thus, mixmaster dynamics may be seen as the dynamics of a single particle
in a time dependent potential.  
The action of Eq. (\ref{theaction}) can be expressed 
by fixing the lapse, $N$, so that
$ N = \sqrt{ {2 \over 3 \pi}} \exp(3 \alpha)$.  This allows us to write 
\cite{MTW},
\begin{equation}
I = \int p_+ d \beta^+ + p_- d \beta^- + p_\alpha d\alpha - {\cal H} d \lambda.
\label{theaction2}
\end{equation}
This action must be supplemented with the condition $ {\cal H} = 0$.
For a fixed value of this potential, the
walls contract as the scale factor increases to the maximum volume slice.
After this slice, the walls of the potential, for a fixed value, recede
back from the center.

To implement the proposed quantization of the theory,
we need to determine the slice with vanishing extrinsic curvature.   
Since all Bianchi type IX classical histories end in  
recollapse \cite{LIN}, we can always identify this slice.
The physical
degrees of freedom are simply the two anisotropies, $\beta^+$ and $\beta^-$. 
In addition, on this slice the volume is maximal.  
Since $\sqrt{h} = e^{3\alpha}$,
to find this maximum volume slice, we must maximize $\alpha$.
From the action of Eq. (\ref{theaction2}) we find,
\begin{equation}
\dot{\alpha}= \{ \alpha, {\cal H} \} = - p_\alpha;
\end{equation}
so that the volume is 
extremized (for finite volume) when $p_\alpha$ vanishes.    
To find the maximum, note that
\begin{equation}
\ddot{\alpha} |_{\dot{\alpha}=0}  =  3 \pi N^2 e^{\alpha} 
{\cal U} (\beta^\pm)
\end{equation}
giving the maximum volume, 
$e^{3 \tilde{\alpha}}$, 
when the potential is less than zero.  This defines an open
region $ \cal R $, 
$ {\cal R} := \{\beta^\pm : {\cal U} (\beta^\pm) < 0 \} $ shown in Fig. 1.
This region is finite \cite{DM}.

The kinematical phase space, $\bar{\Gamma}$, is $R^6$ coordinatized
by $(\alpha, \beta^+ , \beta^- )$ and
conjugate momenta $(p_\alpha, p_+ , p_- )$.  The physical
phase space $\Gamma$ is a four dimensional submanifold
of $\bar{\Gamma}$, which is defined as the quotient of $\bar{\Gamma}$ by
  the solution space of ${\cal H}=0$.  We see that 
\begin{equation}
\Gamma = T^\ast{\cal R}
\end{equation}
coordinatized by $\beta^\pm \in {\cal R}$ and $p_\pm$.  

The coordinates of $\Gamma$ correspond to the
metric and extrinsic curvature of solutions on
the unique maximal slice $p_\alpha=0$.  The quantities 
${\cal O } = \{ \alpha , \beta^\pm , p_\pm \}$
on a slice, say $p_\alpha = \tau$,
could be found by integrating the solutions from $p_\alpha =0$
to $p_\alpha =\tau$ using the Hamilton's equation of
motion (See, for example, \cite{carlo-time}).  We will denote
this set of physical observables ${\cal O} (\tau )$. As all solutions 
end in singularities in some finite
time, the ${\cal O } (\tau )$ are
only defined on a subspace of $\Gamma$ corresponding
to data on the surface $p_\alpha =0$ which specifies non-singular solutions.   

For example, let us consider the observable for volume
of the universe, on the slice where $p_\alpha =\tau $, 
found by solving the Hamiltonian constraint
\begin{equation}
V(\tau ) = e^{3\alpha (\tau )} = \left[   
{  p_+(\tau )^2 + p_-(\tau)^2 - \tau^2 \over 
-{\cal U}(\beta^\pm (\tau ))}
\right]^{3 \over 4}.
\end{equation}
As each of the factors is a complicated function on $\Gamma$,
the explicit form of this can can only be found by integration of the
equation
\begin{equation}
\{ V(p_\alpha ) , {\cal H} \} =0
\end{equation}
on the kinematical phase space $\bar{\Gamma}$
with the initial condition,
\begin{equation}
V(0) = \left (   
{  p_+^2 + p_-^2  \over 
-{\cal U}(\beta^\pm)  }
\right )^{3 \over 4},
\end{equation}
i.e. one must  essentially solve the theory.  At each $\tau$
the subspace $\bar{\Gamma} (\tau)$ defined by $V(\tau ) >0$ is the
domain on which the ${\cal O}(\tau )$ are defined.
That the physical observables are only defined
on subspaces of $\bar{\Gamma}$ is a new feature of cosmology
that raises questions such as whether is the
Poisson bracket of these functions always defined.    Of
great interest is the extent to which this
causes problems for the quantum theory, in either a canonical
or a path integral formalism.  

\begin{figure}
\epsffile{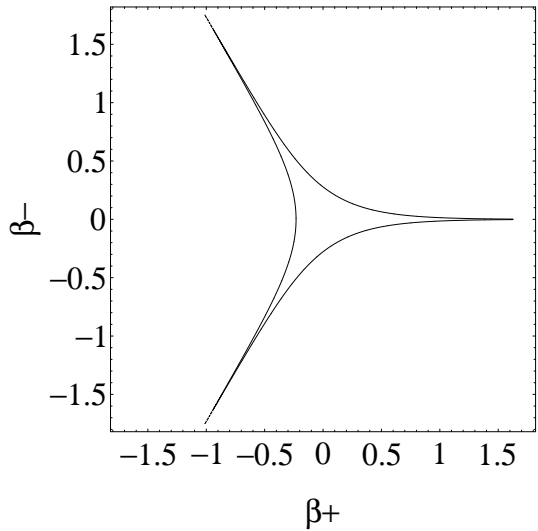}
\caption{The zero contour for the classical potential for Bianchi IX. Three 
``channels,'' 
one along the positive $\beta^-$ axis, and the other two 
sloping diagonally off
in the negative $\beta^-$ direction extend to infinity.  The area of the
interior region, { \cal R}, is finite.}
\end{figure}

\section{Maximal slice canonical quantum theory}

We first quantize the
physical phase space $\Gamma = T^\ast{\cal R}$.  
Since the physical configuration variables, $(\beta^+, \beta^-) \in 
{\cal R}$, and the
conjugate momenta, $(p_+, p_-) \in R^2$, are coordinates on 
the classical phase 
space, take $C^\infty_0$ functions as wavefunctions in the Hilbert space.  
The quantum theory is 
built from the configuration operators $\hat{\beta}^\pm_0$, which act by
multiplication, and the momentum
operators $\hat{p}_\pm^0$ which act by differentiation 
$\hat{p}^0_\pm = -i \hbar \partial_\pm$ on the wavefunctions.  These operators
correspond to measurements of the corresponding quantities
defined on the slice $p_\alpha =0$.  Hence the
subscript and superscript ``$0$.''

We have the inner product,
\begin{equation}
\left\langle \chi | \xi \right\rangle = \int\limits_{\cal R} d^2\beta^\pm_0 \;
 \overline{\chi(\beta^\pm_0)} \; \xi(\beta^\pm_0).
\end{equation}
If the wavefunctions vanish on the boundary of ${ \cal R}$ then
$\hat{\beta}^\pm_0$ and $\hat{p}_\pm^0$ are
Hermitian in this inner product.  

Before considering the problem of observables at times other
than $p_\alpha =0$,  it is interesting to study the operator that
measures the volume on maximal slice.  As this is a property 
of solutions, it is a physical observable, but as it is a function of
data on the maximal slice it can be exhibited explicitly.   It is
convenient to consider the $4/3$ power of the volume, which by Eq. (8) is
\begin{equation}
{V}^{4/3} = { {p}_+^2 + {p}_-^2 \over (- | {\cal U} | )} \equiv q^{ij} 
{p}_i {p}_j,
\end{equation}
written in terms of a metric
defined by $q^{++} = q^{--} = 1/ (-|{\cal U }|)$.  This is non-singular 
in the interior but
blows up on the boundary of $\cal R$.    
Quantum mechanically,  we want to represent ${V}^{4/3}$ as an
hermitian operator.  This can be done by
representing it as the 
Laplace-Beltrami
operator for the metric $q_{ij}$ on the region $\cal R$.  We thus
defined the corresponding operator to be
\begin{equation}
\hat{V}^{4/3} = q^{-1/4} \hat{p}_i q^{ij} q^{1/2} \hat{p}_j q^{-1/4}.
\end{equation}
We note that the form of the inner product Eq. (11) may be interpreted
to mean that the states are half-densities on ${\cal R}$. 

As this is a hermitian operator acting on a finite area, we conjecture that
the volume has a discrete spectra with finite degeneracy.
\footnote{This is delicate.   On the boundaries of the non-compact region the
metric factors, $1/ -|{\cal U }|$, diverge.  The wavefunctions, interpreted as
half-densities, must fall-off faster than the potential.}  This is a familiar
result of one dimensional quantum mechanics.

\section{Evolution in gauge invariant quantization}

In principle, the entire physical quantum theory is contained
in the states and operators on the physical Hilbert space.
However,
to make predictions we need to find 
 observables ${\cal O}^I (\tau )$ for nonzero $\tau$.  If we
were able to do so, all the physical information would be contained
in $N$-time correlation functions of the form
\begin{equation}
\langle \beta^\pm_f |\hat{\cal O}^{1} (\tau_1) \dots
\hat{\cal O}^{N} (\tau_N) 
|\beta^\pm_i \rangle.
\end{equation}
The problem is how to construct these operators given
that we have neither their classical counterparts in closed form
nor the proper subsets of $\Gamma$.  As an example
of such an operator, consider an operator $\hat{V}(\tau )$ corresponding to the
volume of the universe on the slice where $p_\alpha =\tau $.
There are two approaches to this problem:

(1.)  In some appropriate approximation procedure, solve Eqs. (9) and
(10) (or the appropriate equations for other observables) in the classical 
theory, and then, term by term in the approximation, define
an ordering prescription which realizes the classical expression as
a well-defined quantum mechanical operator.    
(We may note the approximation procedure must be able to keep track
consistently of the regions of definition
$\Gamma (\tau )$ which satisfy $\Gamma (\tau_1 ) \in \Gamma (\tau_2 )$
for $\tau_2 > \tau_1$.)

(2.)  Evolve the operators quantum mechanically.  As we have defined 
the quantum theory only for the
physical observables, there is no operator that corresponds to
${\cal H}$; we don't have the quantum mechanical
analogue of Eq. (9).  We can associate a 
non-vanishing Hamiltonian to evolution in a physically meaningful
time variable, and then use that to define an evolution operator.
In the classical theory, this is straightforward.

In the classical phase space, the Hamiltonian which evolves physical 
variables in $p_\alpha $
is the conjugate quantity $\alpha$, which is a time (
$p_\alpha = \tau$) dependent functional on $\Gamma$.
This is given by
\begin{equation}
h (\tau ) \equiv \alpha (\tau ) = {1 \over 4}\ln \left[
p_+(\tau )^2 + p_-(\tau)^2 - \tau^2 \over 
-|{\cal U}(\beta^\pm)|
\right].
\end{equation}
On $\bar{\Gamma}$, this satisfies $\{ h(\tau ) , p_\alpha \} =1$,
so that it generates evolution in $p_\alpha$ on the
constraint surface.  We could construct an evolution operator
in the physical Hilbert space if we
could find the corresponding quantum operator.  However,
this is not easy as we do not know the operators
$\hat{\beta}^\pm (\tau )$ and $\hat{p}_\pm (\tau )$;
these themselves are supposed to be
found by evolution.  In some operator ordering, we would have to find 
operator solutions to the coupled operator equations,
\begin{equation}
\hat{h} (\tau ) = {1 \over 4}\ln \left[
\hat{p}_+(\tau )^2 + \hat{p}_-(\tau)^2 - \tau^2 \over 
-|{\cal U}(\hat{\beta}^\pm)|
\right]
\end{equation}
and
\begin{equation}
{d \hat{\beta}^\pm (\tau ) \over d\tau} = 
[\hat{\beta}^\pm (\tau ) , \hat{h} (\tau ) ]
\end{equation}
\begin{equation}
{d \hat{p}_\pm (\tau ) \over d\tau} = 
[ \hat{p}_\pm (\tau ) , \hat{h} (\tau ) ].
\end{equation}

Needless to say, as the theory cannot be solved at the
classical level, this requires some
approximation procedure.  There are also other issues associated 
with operator ordering
ambiguities such as whether quantum
observables should be hermitian, given that each classical
universe has  a finite lifetime. For, if we find a hermitian
ordering for $\hat{h}(\tau)$ then we ought to be able to evolve
an arbitrary quantum state to arbitrarily large times.  
Given that these practical and conceptual difficulties
face any attempt to proceed with this program, we turn to another
approach to representing evolution based on the path integral.

\section{Path integral representation of physical evolution}

Path integration provides
a practical resolution for these problems.  While the observables 
${\cal O} (\tau )$
cannot be written in closed form on the phase space, they could
be computed by summing over histories. That is, as a
computer averages the effective action over an ensemble of
paths generated by some statistical procedure, it can simply 
go to the points on the path where $p_\alpha =\tau$ and then
tabulate the other observables at those events.  In this
way, averages such as Eq. (14) could be evaluated numerically, with
an ensemble of paths with appropriate
weights or measure. We study one such procedure
in which the path integral measure may
be derived from the physical quantum theory.
In this way, the complementary strengths of the two approaches
may be exploited, for a practical approach to calculating
physical observables non-perturbatively in quantum cosmology.

To construct the path integral corresponding to the
matrix element
\begin{equation}
\langle \beta^\pm_f | {\cal O} (\tau_N ) ...{\cal O} (\tau_1 )
| \beta^\pm_i \rangle
\end{equation}
where $\tau_f > \tau_I > \tau_i$; $I=1,2,3, \dots, N$, 
we express the states in terms of physical
states, elements of ${\cal L}^2 ({\cal R})$.   These are the time evolved kets
\begin{equation}
| \beta_f^\pm \, \tau_f \rangle  = \hat{U}(\tau_i , \tau_f ) |\beta_i^\pm 
\, \tau_i \rangle
\end{equation}
where $\hat{U}(\tau_i , \tau_f )$ is the operator for evolution
$\tau$. 
We derive a path integral expression for Eq.(19) in terms of
a quantum effective action 
$S=\int_{\tau_{i}}^{\tau_{f}}L$, a measure factor $\mu$, and ranges
of integration,
\begin{equation}
\langle \beta^\pm_f | {\cal O} (\tau_N ) ...{\cal O} (\tau_1 )
| \beta^\pm_i \rangle =
{1 \over { \cal N}} \int \left[ d^2\beta d^2 p \mu(\beta,p) \right]
{\cal O} (\tau_N ) ...{\cal O} (\tau_1) 
e^{(i/ \hbar) S}  
\end{equation}
where the ${\cal O} (\tau_I )$ are the classical expressions, expressed as
the value of each observable ${\cal O}$ on the path at the points 
$p_\alpha =\tau $, and 
\begin{equation}
{ \cal N} =\int \left[ d^2\beta d^2 p \mu(\beta,p) ]
\right]e^{(i/ \hbar) S_q} 
\end{equation}
(The brackets a notational device to indicate 
a factor of $1/2\pi$ for each
differential.)
To do this we must first find a path integral expression for
$\hat{U}(\tau_f, \tau_i )$.
The hamiltonian that corresponds to
the choice $\tau = p_\alpha$ [given in Eq. (16)],
may be expressed as 
\begin{equation}
\hat{h}(\tau) = {1 \over 4} 
\left[\ln \left({ \tau^2 \over -|{\cal U}(\hat{\beta}^\pm)|} 
\right )
+ \ln \left( { - \hat{p}_+^2 - \hat{p}_-^2 \over \tau^2} +1 \right) \right].
\end{equation}
The evolution operator is expressed in terms of
the corresponding operator as
\begin{equation}
\hat{U}(0,\tau) = T\exp\left[ {-i \over \hbar} \int_0^\tau \hat{h}(\tau') 
d\tau' \right].
\end{equation}
Without the form of 
$\hat{\beta}^\pm (\tau )$ and $\hat{p}_\pm (\tau )$,  we do have the Hamiltonian $\hat{h}(\tau )$
for finite $\tau$ as an operator on the physical Hilbert space. To 
construct the path integral we make some conjectures
about these operators and their spectra.   These assumptions
cannot be justified directly in the absence of further information 
about the solutions to Eqs. (16), (17), and (18), but they may be
plausible in the light that the resulting path integral agrees
with that constructed by the standard method of gauge fixed
quantization.
We assume that there exist simultaneous
operator solutions to Eqs.(16), (17), and (18) such that they are all hermitian
operators.  As a result time evolution is unitary, so that
the $\hat{\beta}^\pm (\tau )$ and $\hat{p}_\pm (\tau )$
satisfy equal time commutation relations,
$[\hat{\beta}^\pm (\tau ),\hat{p}_{\pm^\prime } (\tau )]=
\delta_{\pm \pm^\prime}$.  
In this case, for each $\tau$, there must be a complete
basis of states made from eigenstates of $\hat{\beta}^\pm (\tau )$,
\begin{equation}
1= \int_{{\cal R}(\tau )} d \beta^\pm \, |\beta^\pm (\tau ) \rangle 
\langle \beta^\pm (\tau ) | 
\end{equation}
where the range of the integral, ${\cal R}(\tau )$, is the
range of the spectra of the operators $\hat{\beta}^\pm (\tau )$.  
In the absence of a construction of the operators, we do not
know this range, but we would like to argue that it is
the whole of $R^2$.  In the classical theory we
have, from the Hamiltonian constraint,
\begin{equation}
{\cal U} (\beta^\pm ) < \tau^2 e^{-4 \alpha}.
\end{equation}
At $\tau=0$, this restricts $\beta^\pm $ to the region $\cal R = {\cal R}(0)$,
However, there are initial classical 
configurations for all values of $p_\alpha$; this means
that for small $\tau$, the left hand side can be arbitrarily large,
which means that there is no limit on how large the anisotropies
$\beta^\pm$ can be at any finite $\tau$.  As a result, if the theory
is to have a good classical limit it seems likely that the spectra
must be unbounded.

Similarly, we assume that there is a complete set of
states
\begin{equation}
1= \int_{R^2} d^2 \beta \, |p_\pm (\tau ) \rangle  \langle p_\pm (\tau ) |. 
\end{equation}
Given all these assumptions we may deduce
\begin{equation}
\langle p_\pm (\tau ) |\beta^\pm (\tau ) \rangle = 
e^{\imath (\beta^+p_+ + \beta^-p_- )}.
\end{equation}
We use standard techniques to construct the path integral.
Inserting complete sets of states in the usual convenient ways, we evaluate
\begin{equation}
h(\tau ) = \langle \beta^\pm (\tau )| \hat{h}(\tau )|p_\pm (\tau ) \rangle
= {1 \over 4} \left [
\ln \left({ \tau^2 \over - |{\cal U}(\beta^\pm )| } \right )
+ \ln \left( { - {p}_+^2 - {p}_-^2 \over \tau^2} +1 \right) \right]
\end{equation}
to find that, 
\begin{equation}
\hat{U}(0,\tau) = \int\left[ 
d^2\beta^\pm d^2p_\pm \right] \exp \left[ 
{ i \over \hbar} \int\limits_0^\tau
p_+ \dot{\beta^+} + p_- \dot{\beta^-} - h(p_\pm,\beta^\pm, \tau) d\tau
\right].
\end{equation}
where the ranges of the integrals are, given our assumptions,
unbounded.  
Comparing to Eq. (21), we then see that in these coordinates the
measure is trivial, the integration regions are unbounded and the
effective lagrangian is given simply by  
$L = p_+ \dot{\beta^+} + p_- \dot{\beta^-} - h(p_\pm,\beta^\pm, \tau)$.

\section{Scale factor time}

To compare this result to the earlier derivation
we'll change from an extrinsic curvature time gauge to a gauge related to the
volume ($\tau = \ln \sqrt{h} = \alpha$).  Geometrically, the measure 
transfers from one slice to another along the orbits
of the gauge transformation.  Unlike $p_\alpha$, $\alpha$ is 
not monotonic on the whole
history of the cosmology.  When the gauge transformation
is made we will choose which half of the history to consider.
Expanding the phase space to include time and the
Hamiltonian, so that
\begin{equation}
\hat{U}(0,\tau) =  \int \left[ d^2\beta^\pm d^2p_\pm d\alpha dp_\alpha
\delta\left( \tau - p_\alpha \right) \delta\left( \alpha - h(\tau) \right)
\right] \exp \left[ {i\over\hbar}  \int p_+\dot{\beta^+}
+ p_-\dot{\beta^-} + p_\alpha \dot{\alpha} d\tau \right]. \label{prop1}
\end{equation}
The new gauge, $\tau=\alpha$, may be introduced as
\begin{equation}
1= \Delta_\alpha \int d\lambda \; \delta\left(\tau - \alpha^\lambda \right)
\end{equation}
where $\alpha^\lambda$ is the gauge transformed $\alpha$. Under small
gauge transformations,
\begin{equation}
\alpha^\lambda = \alpha + \lambda\{\alpha,{ \cal H}\} + O(\lambda^2).
\end{equation}
Thus, $\Delta_\alpha = |\{ \alpha, {\cal H} \}| = |p_\alpha|$. Finally,
we express the delta function involving $\alpha$ in terms of the
Hamiltonian constraint as 
$\delta\left( \alpha - h(\tau) \right)= 
\delta({\cal H}) |\{{\cal H},p_\alpha\}|$. Inserting these 
identities into the propagator of Eq. (\ref{prop1}) gives,
\begin{eqnarray}
\hat{U}(0,\tau) &=&  
\int \left[ d^2\beta^\pm d^2p_\pm d\alpha dp_\alpha d\lambda
\delta\left( \tau - p_\alpha \right) \delta\left( {\cal H} \right)
\delta\left( \tau - \alpha^\lambda \right)
 |\{ \alpha, {\cal H} \}| |\{{\cal H},p_\alpha\}|
\right] \nonumber \\
& & \times \exp \left[ {i\over\hbar} \int p_+\dot{\beta^+} + p_-\dot{\beta^-}
+ p_\alpha \dot{\alpha} d\tau \right].
\end{eqnarray}
Observe that, as before with the new time choice, the extrinsic time choice
satisfies
\begin{equation}
1 = \Delta_{p_\alpha} \int d\lambda \delta \left(\tau - p_\alpha^\lambda
\right).
\end{equation}
By explicitly changing gauge one can check that $\Delta_{p_\alpha}$ is
gauge independent and equals $|\{{\cal H},p_\alpha\}|$ (for small
gauge transformations).  Performing an inverse gauge transformation 
to change gauge we have
\begin{eqnarray}
\hat{U}(0,\tau) & = &  
\int \left[ d^2\beta^\pm d^2p_\pm d\alpha dp_\alpha d\lambda
\delta\left( \tau - p_\alpha^{\lambda^{-1}} \right)
\delta\left( {\cal H} \right) \delta\left( \tau - \alpha^\lambda \right)
|\{{\cal H},\alpha\}| \Delta_{p_\alpha}
\right] \nonumber \\
& & {} \times \exp \left[ {i\over\hbar}\int p_+\dot{\beta^+} 
+ p_-\dot{\beta^-} 
+ p_\alpha \dot{\alpha} d\tau \right] .
\end{eqnarray}
Performing the integration over the gauge parameter, allowing a
delta to eat up the integration over $\alpha$, and exponentiating
the Hamiltonian constraint we find,
\begin{eqnarray}
\hat{U}(0,\tau) & = & \int \left[d^2\beta^\pm d^2p_\pm dp_\alpha |p_\alpha| dN
\right] \nonumber \\
& & \times \exp \left[ {i\over\hbar} \int\limits_0^\tau  p_+ \dot{\beta^+}
+ p_- \dot{\beta^-} + p_\alpha - N { \cal H} d\tau \right] 
\label{prop2}
\end{eqnarray}
the propagator in the physical phase space.  This is the path integral
for the canonical ADM Hamiltonian under the choice $\tau = \alpha$.  With
this choice the Hamiltonian is given by $-p_\alpha$ (the last term in Eq.
(\ref{prop2})) and the lagrange multiplier or lapse is determined so that
$\dot{\alpha} =1$.
The propagator of Eq. (\ref{prop2}) can be integrated to 
a path integral in the phase space $\bar{\Gamma}$;  
as in \cite{SMLS} the integration
over $p_\alpha$ may be performed yielding
\begin{equation}
\hat{U}(0, \tau) = \int\limits_{\bar{\Gamma}} \left[ d^2\beta^\pm d^2p_\pm 
{dN \over N } \right]
\exp \left[ { i \over \hbar } \int\limits_0^\tau  p_+ \dot{\beta^+}
+ p_- \dot{\beta^-} - NH d\tau \right]
\end{equation}
with $H=\left[ p_+^2 + p_-^2 + e^{4\alpha} { \cal U}(\beta^\pm) 
\right]^{1/2}$.
This can be directly compared with the propagator in Section III of 
\cite{SMLS}.   These propagators are identical up to a
measure factor $\mu(\beta^\pm)$.
There is a simple reason for this: in the previous paper \cite{SMLS}
we began with a path integral which included an integration over all
the components of the frame fields and connections, where as here
we specified the theory in Eqs. (1-4) directly in terms of diagonal
gauge.  Thus, the path integral in \cite{SMLS} has an additional
factor in the measure that came from the gauge fixing down to diagonal
gauge.  That factor is non-trivial, as can be seen directly from
Eq. (40) of \cite{SMLS}.  Had we begun the treatment of
\cite{SMLS} with the model defined in diagonal gauge, the results
would have been identical to those found here. 
We conclude that the definition of the path
integral depends on the point at which the reduced theory is taken
to define the quantum theory; reduction does
not commute with quantization.

This measure factor is not relevant for the main question of interest
here, which is whether this ``composite'' approach for defining the physical
theory leads in the end to the same path integral as did the Faddeev-Poppov
ansatz.  The answer is that, at least in this case, if one starts from
the same model, they are the same.

\section{Conclusions}

We have implemented this new approach to quantum cosmology in the restricted,
but non-trivial, 
case of the Bianchi type IX model.  By first, identifying the physical 
degrees of freedom on a slice and then evolving physical observables
off this slice by a sum over
 histories approach, we have shown how it is possible
to learn physics without ``solving the theory.''  We hope that this method
will aid calculation of observables in other models and even the full theory.
Given the physical theory on one slice, physical observables could be
computed numerically.  

This work suggests several directions for further study.  While the technology
appears robust enough to handle the non-linear behavior of Bianchi type IX, 
it would be interesting if a full quantization could be carried out for
one of the more simple Bianchi cosmological models.  This composite approach
also holds some hope for numerical work in that operators at arbitrary times
may be found by computing the propagator for the cosmology.  
Expressed in terms of a path integral, this could
be the basis for a numerical approach to quantum cosmology that has
a physical interpretation grounded in the canonical theory.


\begin{thebibliography}{99}


\bibitem{SMLS} S. Major and L. Smolin, Phys. Rev. D { \bf 51 }, 5475 (1995).

\bibitem{MTW} C. W. Misner, Phys. Rev. {\bf 186}, 1319 (1969), Phys. Rev. Lett.
{\bf 22}, 1071 (1969), or see C. Misner, K. Thorne, and J. Wheeler,
 {\em Gravitation} (Freeman, San Francisco, 1973), pp. 806-814.

\bibitem{BK} V. Belinski, I. Khalatnikov, and E. Lifshitz, Adv. Phys.
{\bf 19 } 525-73 (1973).
 
\bibitem{LIN} X. Lin and R. M. Wald, Phys. Rev. D {\bf 41}, 2444 (1990).
 
\bibitem{KOD}H. Kodama, Prog. Theor. Phys. 80, 1024 (1987) and
Phys. Rev. D 42, 2548 (1990).

\bibitem{MR} V. Moncrief and M. Ryan, Phys. Rev. D {\bf 44} 2375 (1991).

\bibitem{GRA} R. Graham, Phys. Rev. Lett. {\bf 67}, 1381 (1991).

\bibitem{DM} D. Marolf, Class. Quant. Grav. { \bf 12}, 1441-1454 (1995).

\bibitem{relpartpi}J. B. Hartle, K. V. Kucha\u{r}, Phys. Rev. D {\b34},
2323 (1986).

\bibitem{carlo-time}C. Rovelli,  
Phys. Rev. D {\bf 42}, 2638 (1991); {\bf 43}, 442 (1991);
in {\em Conceptual Problems of Quantum Gravity} ed.
A. Ashtekar and J. Stachel,
(Birkhauser,Boston,1991).

\end{thebibliography}
\end{document}